# Robust Interlayer-Coherent Quantum Hall States in Twisted Bilayer Graphene

Dohun Kim[1†], Byungmin Kang[2,3†], Yong-Bin Choi[4], Kenji Watanabe[5], Takashi Taniguchi[6], Gil-Ho Lee[4,7], Gil Young Cho[4,7,8, *], and Youngwook Kim[1, *]

[1]*Department of Physics and Chemistry, Daegu Gyeongbuk Institute of Science and Technology (DGIST), Daegu 42988, Republic of Korea*

[2]*Department of Physics, Massachusetts Institute of Technology, Cambridge, MA, 02139, USA*

[3]*School of Physics, Korea Institute for Advanced Study, Seoul 02455, Republic of Korea*

[4]*Department of Physics, Pohang University of Science and Technology (POSTECH), Pohang 37673, Republic of Korea*

[5]*Research Center for Functional Materials, National Institute for Materials Science, Tsukuba 305-0044, Japan*

[6]*International Center for Materials Nanoarchitectonics, National Institute for Materials Science, Tsukuba 305-0044, Japan*

[7]*Asia Pacific Center for Theoretical Physics, Pohang, 37673, Republic of Korea*

[8]*Center for Artificial Low Dimensional Electronic Systems, Institute for Basic Science (IBS), Pohang 37673, Republic of Korea*

[†]These authors contributed equally

[*]E-mail: gilyoungcho@postech.ac.kr, y.kim@dgist.ac.kr

**Abstract:**

We introduce a novel two-dimensional electronic system with ultrastrong interlayer interactions, namely twisted bilayer graphene with a large twist angle, as an ideal ground for realizing interlayer-coherent excitonic condensates. In these systems, subnanometer atomic separation between the layers allows significant interlayer interactions, while interlayer electron tunneling is geometrically suppressed due to the large twist angle. By fully exploiting these two features we demonstrate that a sequence of odd-integer quantum Hall states with interlayer coherence appears at the second Landau level ($N = 1$). Notably the energy gaps for these states are of order 1 K, which is several orders of magnitude greater than those in GaAs. Furthermore, a variety of quantum Hall phase transitions are observed experimentally. All the experimental observations are largely consistent with our phenomenological model calculations. Hence, we establish that a large twist angle system is an excellent platform for high-temperature excitonic condensation.

**Keywords:** Twisted Bilayer Graphene, Large Twist Angle, Bose-Einstein Condensation, Exciton Condensation, Quantum Hall Effect, Interlayer-Coherence

The interplay between Coulomb interaction and the electronic internal degrees of freedom of materials is a central theme of modern condensed matter physics. Non-trivial topologies permit a range of exotic quantum phases such as topological superconductors [1], integer and fractional Chern insulators [2–8], and the correlated insulating phase [9–17]. A fertile ground to scrutinize emerging phenomena out of electronic correlation and topology is the quantum Hall system [18]. In these systems electronic Coulomb interaction is dominant and often occupies a strongly correlated ground state, where the intricate ordering of the internal degrees of freedom is enriched by the intrinsic topology of quantum Hall states.

For many years, interaction-induced exotic quantum Hall states were only explored in GaAs/AlGaAs heterostructures [19] until similar physics were detected in ZnO/MgZnO [20–22] and bilayer graphene [23–25]. The nature of the quantum Hall effect in monolayer graphene differs to that in the previously mentioned systems. Owing to the unique linear band dispersion entangled with the valley and sublattice degrees of freedom, the electronic wavefunction changes from spinor to scalar depending on the orbital quantum number. Thus, electron–electron interactions at each lowest Landau level (LL) may well differ from those at other levels, and from those of conventional semiconductors. Recent experiments highlight this difference as even-denominator fractional quantum Hall states arise in unusual parameter space [26–28], and charge density waves appear in LLs higher than N = 1 [27, 29].

Another rich area of development in quantum Hall physics is the introduction of layer degrees of freedom, which is realized in double-layer quantum Hall states. The critical parameter here is the interlayer Coulomb interaction, which competes with the intra-layer Coulomb interaction that is controlled by magnetic fields. A unique feature of the double-layer system is an excitonic superfluid that occurs at a half-filled LL in each layer [30–40]. The emergence of the superfluid is determined by the ratio of the interlayer distance, $d$, and magnetic length, $l_b$ (i.e. $d/l_b$). Pioneering works on double-layer systems have revealed that a narrow $d/l_b$ ($d/l_b$ = 1–2) window would introduce excitonic Bose–Einstein condensation. Even though the lower $d/l_b$ limit ($d/l_b < 1$) is believed to support Bose–Einstein condensation and possibly more exotic states, the double-layer community has focused on two active layers separated physically by a few nanometers. Since engineering electrically isolated yet atomically close double layers is challenging, the short-$d$ regime remains relatively unexplored.

In this paper, we introduce a new platform for investigating interlayer-coherent quantum Hall states, namely twisted bilayer graphene with large twist angles. In contrast to conventional bilayer quantum Hall systems, our system enables ultrastrong interlayer interaction owing to the atomically thin spacing between the layers. For instance, the ratio between the interlayer distance and magnetic length is exceedingly small, $d/l_b \sim O(10^{-2}) \ll 1$. Furthermore, we employ a large twist angle to suppress single-electron tunneling between the layers. Note that when the two layers are separated by such a short distance, interlayer electronic hopping is often significant, and the resulting band structure can be vastly different from its single-layer counterpart. Concomitantly, the layer degree of freedom is no longer relevant owing to strong hybridization, as happens in the magic angle twisted bilayer graphene [9–17]. However, for sufficiently large twist angles ($\theta > 3°$), the geometric constraints make it difficult for electrons to hop between the two layers [41], which results in near-intact layer degrees of freedom and band structures. These two novel ingredients, the extremely large interlayer coupling and suppression of single-electron tunneling, make our system an ideal avenue for exploring the strong-correlation physics of bilayer quantum Hall states. Indeed, we find distinctive signatures of odd-integer quantum Hall states and various phase transitions by experiment. Our experimental results are further corroborated by theoretical calculations.

We fabricated large-twist-angle bilayer graphene using a pick-up transfer technique. We measured three devices in total, and all devices showed qualitatively and quantitatively similar results. In the main text we present Device 1 while magnetotransport results for Devices 2 and 3 are discussed in the Supporting Information. Figure 1a shows longitudinal resistance as a function of the top and bottom gate voltages, supplied by few-layer graphene layers with a thickness of approximately 10 nm. The top and bottom gate dielectrics were fabricated using 45 and 33 nm of h-BN, respectively. Here, the target twist angle between the two graphene layers was 10° or larger. We used a standard cut-and-stack technique [42] or stack two different graphene monolayers to assemble our van der Waals heterostructure. However, we could not use scanning probe microscopy [43] or Raman spectroscopy [44] to identify the twist angle of our device as the twisted bilayer graphene was fully encapsulated by h-BN and graphite layers. Transport experiments are an effective tool to determine the twist angle, but interlayer signatures can only resolve small angles. Small twist angles can be obtained from the position of the secondary Dirac point and Landau fan diagram. However, no specific features are available for large-angle systems, as is the case here. Without a magnetic field, we do not see any hint of superlattice formation up to a density of approximately $10^{13}$ cm$^{-2}$, which is the position of the secondary Dirac point for a 2° twist angle, and of the van Hove singularity point for a 3° twist angle. This indicates that our device has twist angles larger than 3°.

Magnetotransport further indicates that the twist angle exceeds 3°. Figure 1b shows the results a dual-gate sweep at $B = 1$ T and $T = 1.6$ K. Twisted bilayer graphene inherits properties from monolayer graphene, such as spin and valley degrees of freedom. In addition, two monolayers cause an additional layer degree of freedom. Therefore, the LL in twisted bilayer graphene has an 8-fold degeneracy (when the Zeeman energy is ignored). When a finite displacement electric field, $D/\varepsilon_0$, is applied to the top and bottom graphene layers, the layer degrees of freedom is lifted. Thus, multiple quantum Hall transitions in $R_{xx}$ are depicted in Figure 1b. Magnetotransport results do not show any signatures of van Hove singularity or quantum Hall states from other symmetric points. From the zero-field and $B$-field datasets, we can conclude that our device has a sufficiently large twist angle. Furthermore, measurements based on optical images of the three devices indicate that our twist angle is approximately 10°, as intended (see supporting information S1).

To investigate the quantum Hall states more clearly, we unload our device from a He$^4$ cryostat and loaded it into a dilution refrigerator. The quantum Hall effect at the N = 0 LL recorded at maximum magnetic field ($B = 15$ T) and base temperature ($T = 30$ mK) is shown in Figure 2a. Conventional two-flux composite fermion states emerge at filling, for example $\nu_{tot} = \nu_{top} + \nu_{bottom} = 2 + 1/3, 2 + 2/5, 2 + 3/5, 2 + 2/3$ [27, 28]. We found several quantum phase transitions tuned by $D/\varepsilon_0$ at $\nu_{tot} = 1, 2,$ and 3 as a function of $D/\varepsilon_0$, but did not find stable interlayer coherence quantum Hall states at $D/\varepsilon_0 = 0$ as shown in Fig 2b. We observe odd integer quantum Hall states only when displacement electric field is applied across the layers as shown in Fig 2c. This contrasts with our previous research on a 2° device, where stabilized $\nu_{tot} = 1$ and 3 states at $T = 1.3$ K for $D/\varepsilon_0 = 0$ were found owing to the larger interlayer correlations for the smaller twist angle [45]. For the 2° device, the higher odd-integer filling states were not accessible because of the secondary charge neutrality points and van Hove singularities, which originate from the strong interlayer hybridization [46]. (Further discussion between the 2° and large-angle devices are referred to the Supporting Information)

Figure 3a shows the longitudinal resistance of the N = 1 LL as a function of the total filling factor and displacement electric field in 15 T magnetic field at 30 mK. Owing to the high quality of our device, we could access the fractional quantum Hall states including the two-flux composite fermion states in the N = 1 LL. However, the paired states of composite fermions such as even-denominator states are missing, similar to the N = 1 LL of monolayer graphene.

We found several triangular high-resistivity regions separated by low-resistivity lines. Along the low-resistivity lines (denoted by blue (orange) lines in Figure 3d) the filling factor of the top (bottom) layer is approximately tuned to an integer. Along this line most charge carriers in the top (bottom) layer are expected to form an integer quantum Hall-like state, which can result in low resistivity. On the other hand, the bilayer system as a whole fail to form a stable quantum Hall phase owing to the mismatch in the total filling factor. Hence, stable quantum Hall phases can emerge only where the blue and orange lines cross, i.e. when the total filling is tuned to an integer.

At a fixed-integer total filling factor different quantum Hall phases are realized as a function of $D/\varepsilon_0$, where each phase is characterized by different layer polarization. Numerical results from our theoretical model in Figure 3d and Figures 4a–4c suggest the nature of the observed states (for the details see section II in the supporting information). For example, at $\nu_{tot} = 8$ canted anti-ferromagnetic phases are realized in the top and bottom layer (Figure 4a), which is stable over a finite $D/\varepsilon_0$ window [47]. For a larger $D/\varepsilon_0$, there is a phase transition toward a state with nonzero layer polarization (Figure 4b). In this state, the layer polarization varies continuously as a function of $D/\varepsilon_0$. As $D/\varepsilon_0$ increases there is a further transition to a fully layer-polarized quantum Hall phase (Figure 4c).

Notably, a sequence of odd-integer quantum Hall phases emerges at $D/\varepsilon_0 = 0$, which clearly lies beyond simple single-particle physics. $\nu_{tot} = 5, 7, 9,$ and $11$ are displayed in Figure 3a, 3b, and 3c. Because of the valley and layer degeneracy of each LL, non-interacting electrons can give rise to even-integer quantum Hall states when $D/\varepsilon_0 = 0$. These interaction-enabled odd-integer quantum Hall states become more distinctive in our transport data when the magnetic field is increased from 9 T to 15 T.

In our device, intra-layer and interlayer Coulomb interaction are distinct sources for correlation. From the intra-layer interactions, the combination of two (almost decoupled) fractional quantum Hall states from the top and bottom layers can develop an odd-integer quantum Hall state. There are two possible scenarios: either two identical even-denominator fractional quantum Hall states, i.e. $1/2 + 1/2$, or two odd-denominator fractional quantum Hall states with different numerators, e.g. $1/3 + 2/3$. However, neither is consistent with our experiments. First, we maintained a zero-displacement electric field, so an electronic density imbalance between the top and bottom layers seems unlikely. Thus, different filling fractions in two layers can be discounted. Second, even-denominator fractional quantum Hall states do not emerge in the $N = 1$ LL of monolayer graphene [26–29]. Therefore, the even-denominator state is unlikely to be a good basis for explaining the odd-integer quantum Hall states in our device. Having excluded these possibilities, we suggest that interlayer interactions explain the odd-integer quantum Hall states.

We used Hartree–Fock calculations to find the ground states at each odd-integer filling, which were indeed characterized by interlayer coherence as illustrated in Figures 4d and 4e. The layer polarization continuously increases as $D/\varepsilon_0$ increases, so the system retains its gap with nonzero interlayer coherence. This is because an interlayer-coherent bilayer quantum Hall state is expected to be robust against the charge imbalance between layers, as in GaAs [48–50]. Such smooth evolution of the odd integer quantum Hall states under the increment of $D/\varepsilon_0$ has been also observed in other double layer quantum Hall systems [37–40]. Upon further increasing $D/\varepsilon_0$, there is a phase transition to a fully layer-polarized phase with the ferromagnetic spin order. Similar evolution of the states is expected for the even-integer quantum Hall states (see supporting information S4).

Through observation of thermally excited behavior, we could extract the size of the excitation gap. Figures 5a and 5b show the temperature dependence of the quantum Hall curves. The temperature in the dilution refrigerator was increased from 30 to 700 mK and higher temperature data were recorded in a He$^4$ cryostat. While transferring the device, we noticed a degradation in the transport quality, and hence the quantum Hall states were not as strong in the second dataset. Figure 5b depicts the temperature-dependent resistivity at 9 T and $D/\varepsilon_0 = 0$. At $T = 1.6$ K, all integer quantum Hall states and resistivity minima at odd-integer fillings are developed and show thermally activated behavior as temperature increases. Figure 5c shows quantum Hall gaps obtained experimentally. Unlike the mK-width of the quantum Hall gap in the weak-coupling regime, the gap for the four resistivity minima at odd-integer fillings is approximately 1 K, reflecting the strong interlayer coupling. In comparison to the odd-integer quantum Hall states, the size of the quantum Hall gap for even-integer quantum Hall states is approximately 3–4 K. The largest gap is $\nu_{tot}$ = 8, and $\nu_{tot}$ = 6 and 10 are slightly smaller. Quantitatively, these agree well with the calculated gap sizes depicted in Figure 5d. We note that the theoretical gaps are larger than the ones from experiments, which could be attributed to the Hartree–Fock calculation, which often underestimates the quantum fluctuations [51–53]. However, the even–odd effects are evident, as the gaps at even fillings are larger than those at odd fillings. This is because of the hierarchy of the interlayer interactions and intralayer interactions. (See supporting information S4)

Interestingly, the stable odd-integer quantum Hall phases are only observed in the N = 1 LL. At first, this seems at odds with conventional wisdom; the quantum Hall phases tend to be more stable in the N = 0 LL than in the N = 1 LL. While the electron–hole excitation gap generally decreases as N increases, the energy gap of nonlocal pseudospin excitations such as skyrmions may be smallest at N = 0 LL, as discussed by Shi et al. [54]. Furthermore, in the N = 0 LL the gap becomes comparable to or smaller than that of the particle–hole excitations [54, 55] (see supporting information S2). This is in accordance with the experimental observations of Shi et al. [54], in which odd-integer quantum Hall phases were also absent in the N = 0 LL. Indeed, by applying much higher magnetic fields (B > 15 T) to Device 2, we found weak but still discernible signals for the odd-integer quantum Hall states in the N = 0 LL, with a smaller gap than those at the N = 1 LL (see supporting information S3).

In conclusion, we have investigated the quantum Hall states in high-quality twisted bilayer graphene with large twist angles. Magnetoresistance and temperature-dependent transport measurements together with numerical calculation revealed that the interlayer-coherent states emerge as the ground states at odd-integer fillings. By carefully adjusting the displacement electric fields, we observed various integer quantum Hall states where the intricate ordering of the internal valley, spin, and layer degrees of freedom was expected, and the transitions between these states. We determined the size of the quantum Hall gaps using temperature-dependent measurements and found that the gaps for interlayer-coherent states were approximately 1 K, which is one or two orders of magnitude larger than conventional GaAs double layers, which we attribute to the ultrastrong interlayer interactions. We conclude that twisted bilayer graphene with large twist angles provides an excellent avenue for interaction-enabled interlayer-coherent quantum Hall states that are more robust than their counterparts in conventional double layer two-dimensional electron gas systems.

## Associated Contents

### Supporting Information

The Supporting Information is available free of charge at

Twist angle and additional transport data, Odd-integer quantum Hall states in the $N = 0$ LL & Discussion between low and large twist angle devices, The Hatree-Fock method, Mean-field solutions, Skymion-anti skymion excitation.

## Author Information

### Corresponding Authors

Gil Young Cho - Department of Physics, Pohang University of Science and Technology (POSTECH), Pohang 37673, Republic of Korea; Asia Pacific Center for Theoretical Physics, Pohang, 37673, Republic of Korea; Center for Artificial Low Dimensional Electronic Systems, Institute for Basic Science (IBS), Pohang 37673, Republic of Korea;

Email: gilyoungcho@postech.ac.kr

Youngwook Kim - Department of Physics and Chemistry, Daegu Gyeongbuk Institute of Science and Technology (DGIST), Daegu 42988, Republic of Korea

Email: y.kim@dgist.ac.kr

### Authors

Dohun Kim - Department of Physics and Chemistry, Daegu Gyeongbuk Institute of Science and Technology (DGIST), Daegu 42988, Republic of Korea

Byungmin Kang - Department of Physics, Massachusetts Institute of Technology, Cambridge,

MA, 02139, USA; School of Physics, Korea Institute for Advanced Study, Seoul 02455, Republic of Korea

Yong-Bin Choi - Department of Physics, Pohang University of Science and Technology (POSTECH), Pohang 37673, Republic of Korea

Kenji Watanabe - Research Center for Functional Materials, National Institute for Materials Science, Tsukuba 305-0044, Japan

Takashi Taniguchi - International Center for Materials Nanoarchitectonics, National Institute for Materials Science, Tsukuba 305-0044, Japan

Gil-Ho Lee - Department of Physics, Pohang University of Science and Technology (POSTECH), Pohang 37673, Republic of Korea; Asia Pacific Center for Theoretical Physics, Pohang, 37673, Republic of Korea

**Author contributions**

Y. K and G.Y.C conceived the project. D. K carried out the device fabrication and performed the low-temperature measurement with Y-B.C and G-H.L. The theory was performed by B. K and G. Y. C. T. T and K. W synthesized the h-BN crystals. All authors contributed to the manuscript writing.

**Acknowledgments**

We thank Klaus von Klitzing, Jurgen H. Smet, Joseph Falson, Ding Zhang, Kwon Park, and Jun Sung Kim for fruitful discussions. This work has been supported by the Basic Science Research Program NRF-2020R1C1C1006914 through the National Research Foundation of Korea (NRF) and by the DGIST R&D program (22-CoE-NT-01) of the Korean Ministry of Science and ICT. This research was also supported by BrainLink program funded by the Ministry of Science and ICT through the National Research Foundation of Korea(2022H1D3A3A01077468). K.W. and T.T. acknowledge support from JSPS KAKENHI (Grant Numbers 19H05790, 20H00354 and 21H05233)　and A3 Foresight by JSPS. G.Y.C. is supported by the NRF of Korea (Grant No. 2020R1C1C1006048) funded by the Korean Government (MSIT) as well as the Institute of Basic Science under project code IBS-R014-D1. G.Y.C. is also supported by the Air Force Office of Scientific Research under Award No. FA2386-20-1-4029 and No. FA2386-22-1-4061. G.Y.C acknowledges Samsung Science and Technology Foundation under Project Number SSTF-BA2002-05. B.K. acknowledges the support by DOE office of Basic Sciences Grant No. DE-FG02-03ER46076, KIAS individual Grant PG069402 at Korea Institute for Advanced Study, and the National Research Foundation of Korea (NRF) grant funded by the Korea government (MSIT) (No. 2020R1F1A1075569). Y.-B.C. and G.-H.L. were supported by the MSIT (Ministry of Science and ICT), Korea, under the ITRC (Information Technology Research Center) support program (IITP-2022-RS-2022-00164799) supervised by the IITP (Institute for Information & Communications Technology Planning & Evaluation), under Basic Science Research Institute Fund (2021R1A6A1A10042944), and under National Research Foundation of Korea (2022M3H4A1A04074153).

**Competing interests**

The authors declare no competing interest.

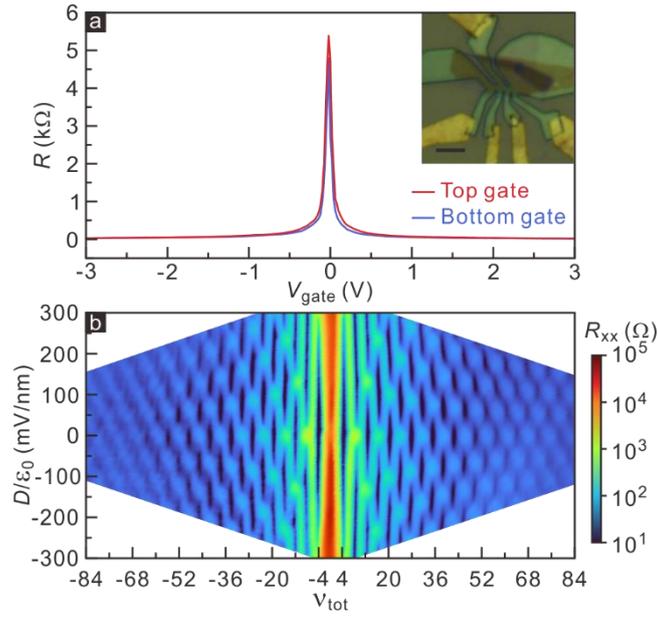

**Figure 1| Transport properties of large-angle twisted bilayer graphene.** (a) Resistance as a function of gate voltage at a temperature of 1.6 K. Red and blue solid lines represent the top and bottom gate voltage dependence, respectively. The inset shows an optical microscope image of the twisted bilayer graphene device. The scale bar is 5 μm. (b) Color map of longitudinal resistance, $R_{xx}$, in the plane described by the total filling factor and the displacement field at $B = 1$ T and $T = 1.6$ K.

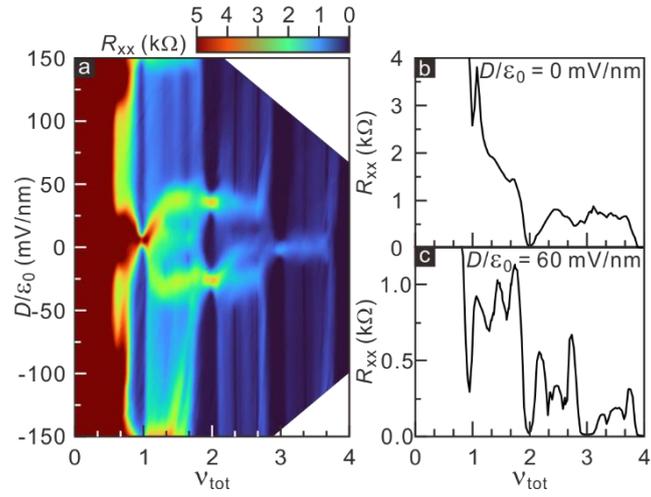

**Figure 2| The quantum Hall effect in the N = 0 LL.** (a) A contour map of $R_{xx}$ of the N = 0 LL in the ($\nu_{tot}$, $D/\varepsilon_0$) plane at $B$ = 15 T and $T$ = 30 mK. (b) Longitudinal resistance as a function of the total filling factor with zero displacement field. (c) The same as (b) but under a 60 mV nm$^{-1}$ displacement field.

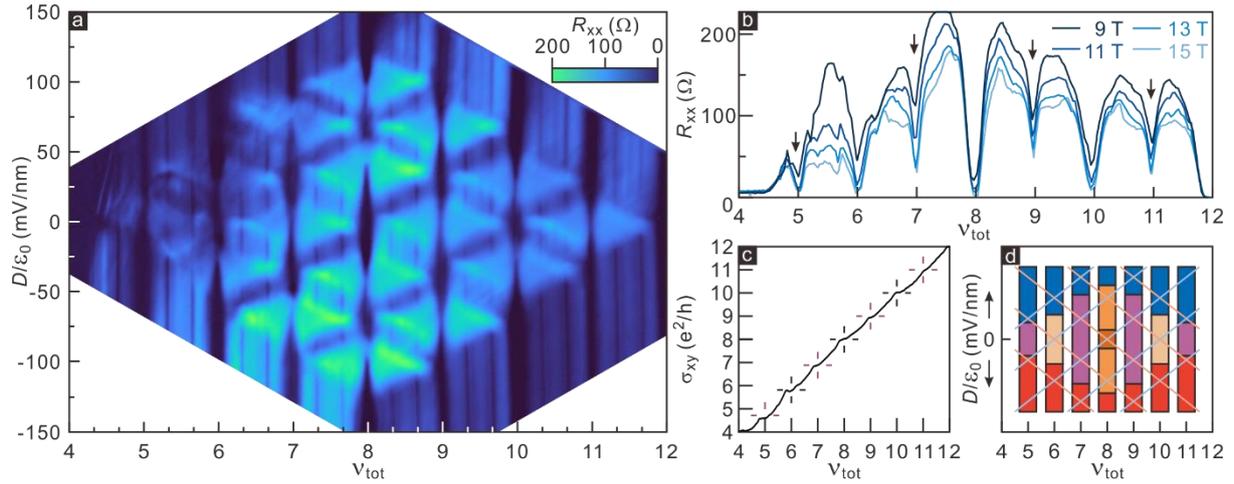

**Figure 3| A cascade of quantum Hall states in the N = 1 LL.** (a) Contour map of $R_{xx}$ as a function of the displacement field and $\nu_{tot}$ for the N = 1 LL at $B$ = 15 T and $T$ = 30 mK. (b) Longitudinal resistance as a function of total filling factor under a fixed magnetic field from 9 to 15 T in 2 T increments in the absence of a displacement field. Black arrows highlight resistivity minima at the odd-integer filling factors. All data were recorded at $T$ = 30 mK (c) $\sigma_{xy}$ as a function of $\nu_{tot}$ at $B$ = 9 T and $T$ = 1.6 K. Black and purple solid lines denote even and odd quantized Hall plateaus, respectively. (d) Theoretical quantum phase diagram of the N = 1 LL crossing as a function of the displacement field and total filling factor. Blue (orange) solid lines show the top (bottom) graphene layer filling factor. Layer coherence phases at the odd-integer quantum Hall states are shown by purple boxes, whereas orange-shaded boxes at even filling display canted anti-ferromagnetic states and valley pairing states. Blue and red boxes represent fully layer-polarized quantum Hall phases.

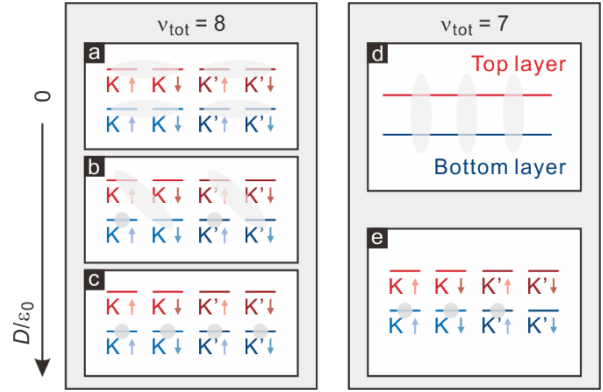

**Figure 4| Evolution of quantum Hall phases.** Schematics of quantum Hall phases of $\nu_{tot} = 8$ (a)–(c) and $\nu_{tot} = 7$ (d), (e) as the displacement field increases from 0. The displacement electric field is shown on the vertical axis. Layer degrees of freedom are shown in blue (bottom layer) and red (top layer). The valley and spin indices are denoted using text (K, K') and arrows with the corresponding layer colors. Light-grey ellipses represent single-particle electron states with a coherent superposition of two symmetry-broken states and layer-polarized states are indicated by dark-grey circles. (a) The canted antiferromagnetic phase is realized in each layer. (b) The valley-coherence antiferromagnetic phase with partial layer polarization. In this phase, layer polarization changes continuously as a function of the displacement field. (c) The fully layer-polarized phase. All the N = 1 electron states in the bottom (top) layer are occupied (emptied). (d) The layer-coherent phase under a small displacement field. This layer-coherent phase with mixed valley and spin degrees of freedom is stable to a finite displacement field with varying layer polarization as a function of the displacement field. (e) The fully layer-polarized phase.

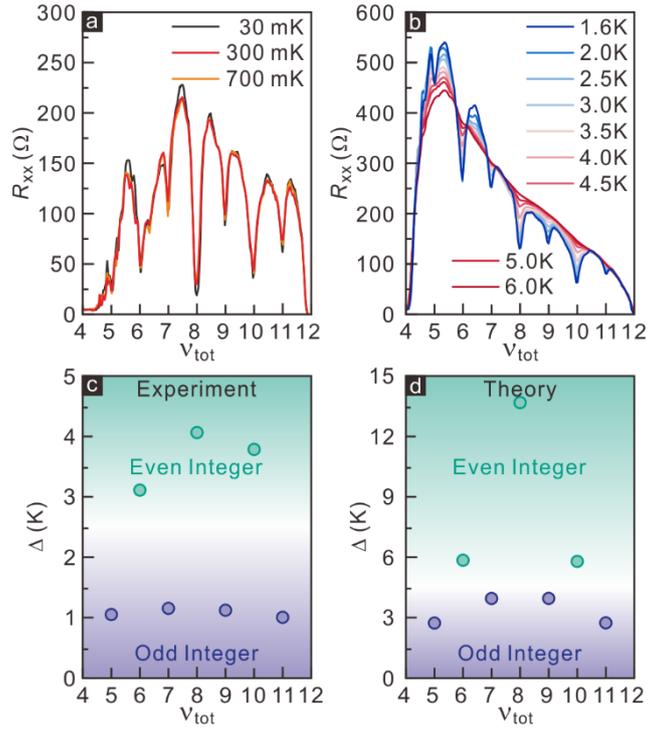

**Figure 5| Thermally excited quantum Hall states and energy gap in the N = 1 LL at 9 T.** (a) Temperature dependence of $R_{xx}$ plotted against the total filling factor without a displacement field at $T$ = 30 mK, 300 mK, and 700 mK. (b) The same as (a) but for $T$ = 1.6 K to 6.0 K. The line color indicates different temperatures, with reference to the key. (c) Experimentally determined energy gap ($\Delta$) of the even- and odd-integer quantum Hall states. (d) The theoretical energy gap of even- and odd-integer quantum Hall states calculated using the Hartree–Fock method.

# Supporting Information on "**Robust Interlayer-Coherent Quantum Hall States in Twisted Bilayer Graphene**"

Dohun Kim[1†], Byungmin Kang[2,3†], Yong-Bin Choi[4], Kenji Watanabe[5], Takashi Taniguchi[6], Gil-Ho Lee[4,7], Gil Young Cho[4,7,8, *], and Youngwook Kim[1, *]

*[1]Department of Physics and Chemistry, Daegu Gyeongbuk Institute of Science and Technology (DGIST), Daegu 42988, Republic of Korea*

*[2]Department of Physics, Massachusetts Institute of Technology, Cambridge, MA, 02139, USA*

*[3]School of Physics, Korea Institute for Advanced Study, Seoul 02455, Republic of Korea*

*[4]Department of Physics, Pohang University of Science and Technology (POSTECH), Pohang 37673, Republic of Korea*

*[5]Research Center for Functional Materials, National Institute for Materials Science, Tsukuba 305-0044, Japan*

*[6]International Center for Materials Nanoarchitectonics, National Institute for Materials Science, Tsukuba 305-0044, Japan*

*[7]Asia Pacific Center for Theoretical Physics, Pohang, 37673, Republic of Korea*

*[8]Center for Artificial Low Dimensional Electronic Systems, Institute for Basic Science (IBS), Pohang 37673, Republic of Korea*

[†]These authors contributed equally

[*]E-mail: gilyoungcho@postech.ac.kr, y.kim@dgist.ac.kr

**Methods**

**Device fabrication**

We used an Elvacite stamp based on the van der Waals pick-up transfer method to assemble a graphite/h-BN/twisted bilayer graphene/h-BN/graphite/h-BN (bottom to top) heterostructure, where two graphite layers were responsible for the gate voltage knob, and the thickness of the h-BN layers was between 30 and 50 nm.

To increase the bubble-free area, the device was annealed at 600°C in a forming gas (Ar/H$_2$) environment for approximately 30 minutes. Electrodes were patterned using an electron-beam lithography technique involving a 600-nm-thick double-layer PMMA resist coating. The Hall bar geometry and edge contact were defined using reactive ion etching (RIE) with a mixture of CF$_4$ (40 sccm) and O$_2$ (4 sccm). The RIE chamber pressure was $3 \times 10^{-1}$ torr during treatment and we exposed the device for 40 seconds with a power of 40 W. Then, 70-nm-thick Au electrodes were deposited on top of a 10-nm-thick Cr adhesion layer in an electron-beam evaporator chamber with a base pressure of $5 \times 10^{-7}$ torr.

**Transport measurement**

We employed a standard lock-in technique with $I = 100$ nA and $f = 13.333$ Hz for the twisted bilayer graphene device. Magnetotransport measurements were taken in two different cryostats; a He$^4$ dry cryostat (JANIS) and a Leiden dilution refrigerator with a base temperature of 30 mK.

The displacement electric field $D/\varepsilon_0$ was calculated as $D/\varepsilon_0 = (C_\mathrm{T}V_\mathrm{TG} - C_\mathrm{B}V_\mathrm{BG})/2\varepsilon_0$, where $C_\mathrm{T(B)}$ and $V_\mathrm{TG(BG)}$ are the capacitance per unit area of the h-BN dielectric layer, the top and bottom gate voltages are $V_\mathrm{TG}$ and $V_\mathrm{BG}$, respectively, and $\varepsilon_0$ is the vacuum permittivity. $C_\mathrm{T(B)}$ and $V_\mathrm{TG(BG)}$ can be used to calculate the total filling factor according to $\nu_\mathrm{tot} = n_\mathrm{tot}h/eB$, where $n_\mathrm{tot}$ is the total density given by $n_\mathrm{tot} = (C_\mathrm{T}V_\mathrm{TG} + C_\mathrm{B}V_\mathrm{BG})/e$. Finally, $e$ is the elementary charge, $h$ is the Planck constant, and $B$ is the magnetic field.

# I. Experimental Data

## S1. Twist angle and additional transport data

Our target twist angle is approximately 10° because it is far from the magic angle, and the Fermi velocity of the 10° twist device is almost identical to that of single-layer graphene [1]. The graphene and its final heterostructure (before electron-beam lithography) are pictured in Figure S1. Devices 1 and 3 were fabricated using the cut-and-stack method, and two different graphene sheets were used for Device 2. We can determine the twist angle of our devices by measuring two graphene edges in the final heterostructure image. The twist angle was measured in the range of 10–12°. There is some uncertainty [2]; however, a 10° twist angle with a 1–2° error has no significant bearing on our experiment and conclusion.

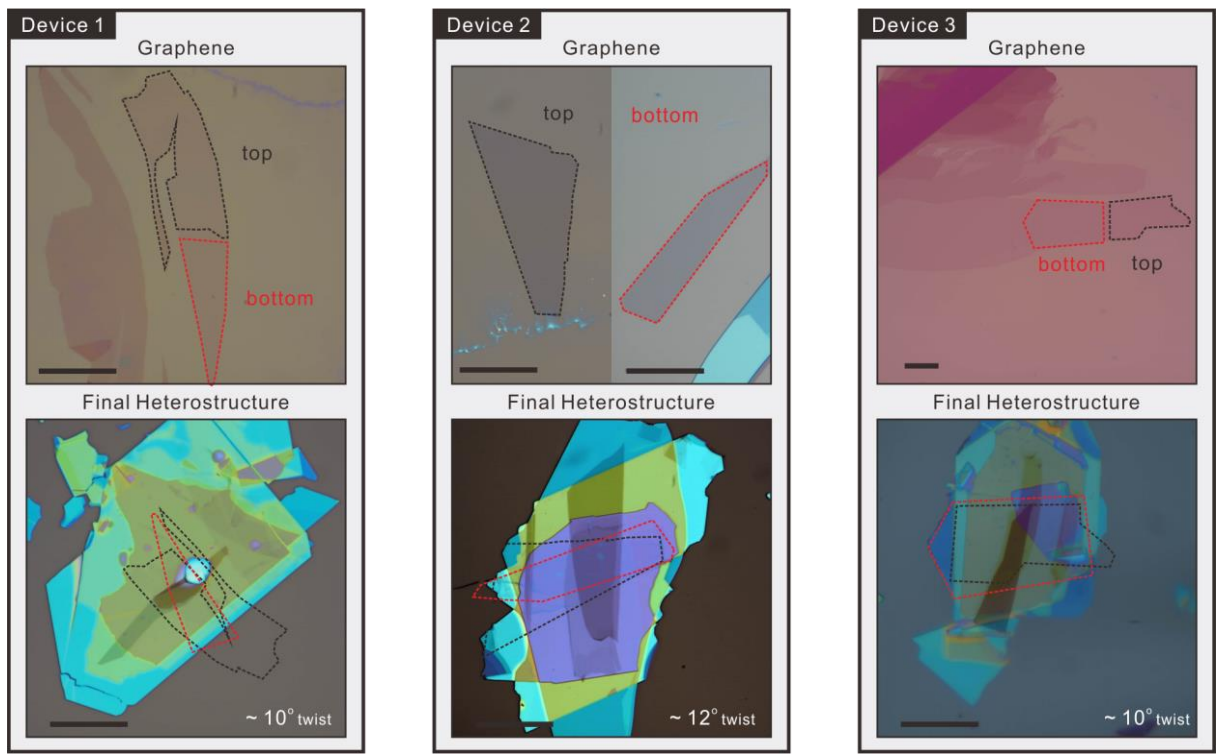

**Figure S1| Optical images of graphene and the final heterostructure of Devices 1, 2, and 3.** The black (red) dotted-line indicates the top (bottom) graphene. The scale bar is 10 μm.

Figure S2b shows single line trance of longitudinal resistivity for different displacement electric fields at $T = 30$ mK and $B = 15$ T. Here we selected a few values of electric fields for the plot, where the quantum Hall transitions appear. They are also indicated by color arrows. As we discussed in main text, all quantum Hall states are smoothly developed with changing $D/\varepsilon_0$,

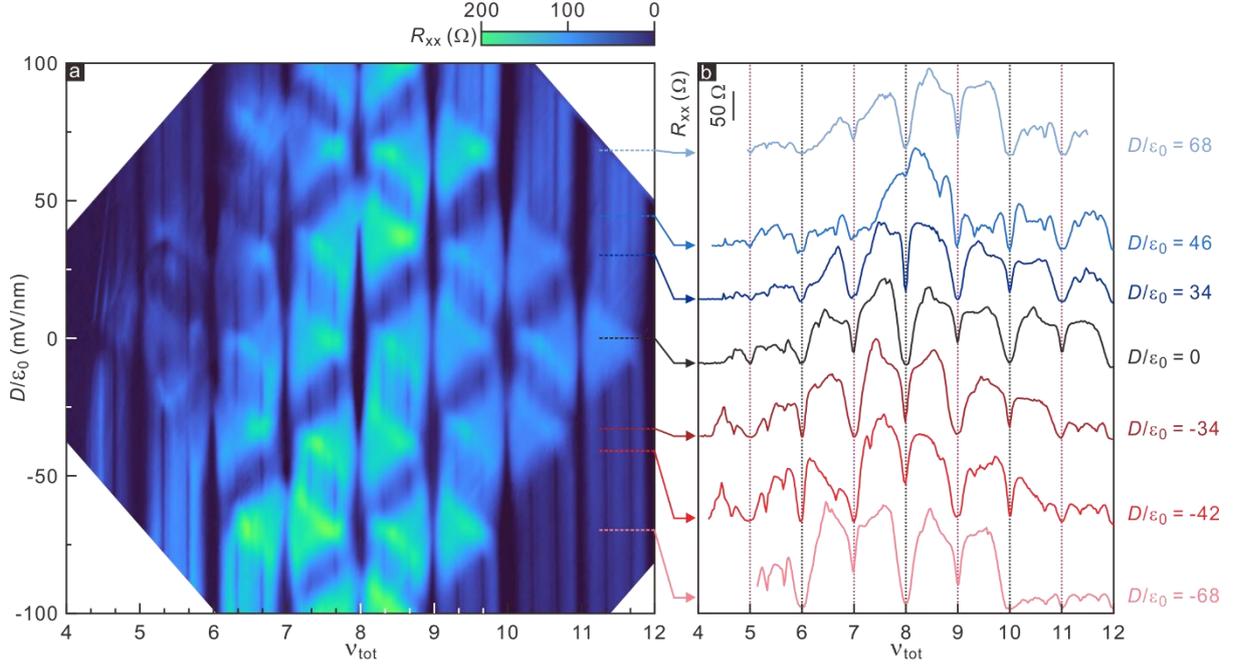

**Figure S2| The quantum Hall curves on Devices 1.** (a) The same plot as Fig3(a) but smaller displacement window. (b) Trances of $R_{xx}$ for different $D/\varepsilon_0$. The scale bar is 50 Ω. Line trances are vertically offset for clarity. Different colors indicate different $D/\varepsilon_0$. These values of the displacement electric fields are denoted on the right panel of Fig S2b. In addition, connecting color arrows from Fig. S2a to Fig. S2b also show the displacement electric field for each line trace.

Figure S3 presents magnetotransport data at the N = 1 LL in Devices 2 and 3. Both devices show similar odd-integer quantum Hall states and transitions. In the main text, we focused on the electron side of Device 1 as the hole side showed relatively weak transport signals. However, the hole regime of Devices 2 and 3 performed better than the electron regime. Therefore, we discuss the hole dataset in this supporting information; although electron and hole regimes, in principle, obey the same physics. Plateaus are found at $D/\varepsilon_0 = 0$ and zero $R_{xx}$ resistance, as shown in Figure S1b. All features were weaker in Device 3 than Device 2, as shown in Figure S1d. These effects were mostly a result of the measurement conditions, i.e. the temperature and magnetic field. Device 2 was measured in a top-loading dilution refrigerator with a 30 mK base temperature, while Device 3 was measured in a closed-cycle He[4] cryostat, with base temperature of 1.6 K. Magnetic fields also differed slightly, and both measurements were conducted with fixed magnetic fields of 10 and 9 T, respectively.

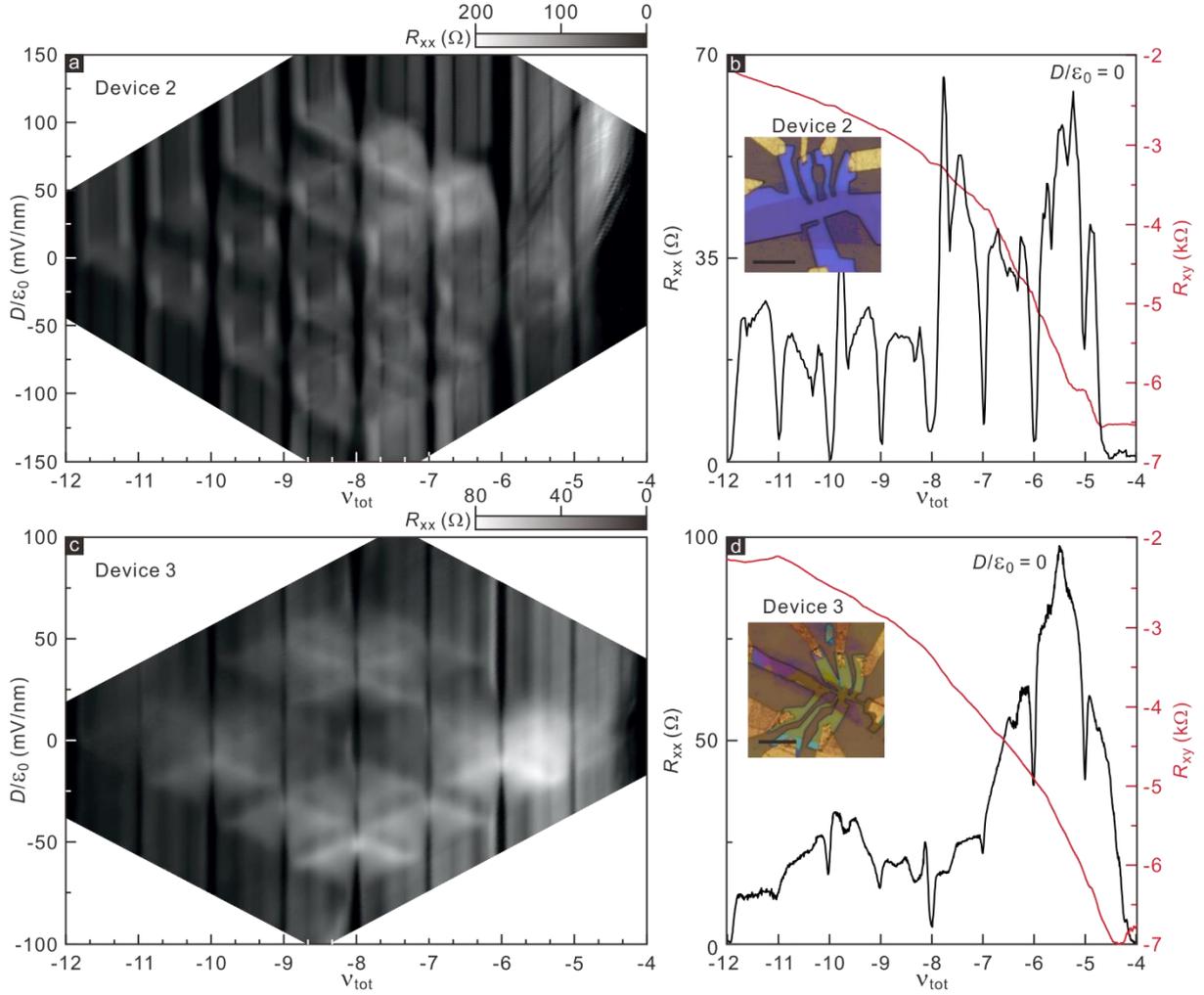

**Figure S3| The magnetotransport properties of Devices 2 and 3.** (a) Greyscale rendition of $R_{xx}$ as a function of total filling factor and displacement electric field for Device 2 at $B = 10$ T and $T = 30$ mK. (b) $R_{xx}$ (black) and $R_{xy}$ (red) curves as a function of $\nu_{tot}$ at $D/\varepsilon_0 = 0$. Inset shows an optical image of Device 2 with a 10 μm scale bar. (c), (d) The same plots as (a) and (b) but for Device 3. Data for Device 3 were recorded at $T = 1.6$ K and $B = 9$ T. The inset in (d) shows an optical image of Device 3 with a 10 μm scale bar.

## S2. Odd-integer quantum Hall states in the N = 0 LL & Discussion between low and large twist angle devices

We found odd-integer quantum Hall states at the N = 0 LL in Device 2. Figure S4a shows a colormap of the longitudinal resistance at $T = 30$ mK and $B = 19$ T. The odd-integer quantum Hall states at $\nu_{tot} = -1$ and $-3$ are well developed, and they are very stable under varying displacement electric fields; a similar effect was observed at the N = 1 LL. However, the $\nu_{tot} = -1$ and $-3$ states at $D/\varepsilon_0 = 0$ require a larger magnetic field than those of the odd-integer quantum Hall states at the N = 1 LL. Figures S2b–2f show single-line traces of $R_{xx}$ at odd-integer filling for the N = 0 and 1 LLs along the displacement electric field axis. The five lines represent five different magnetic fields: 8 T, 10 T, 13 T, 15 T, and 19 T. We found a resistance spike at $D/\varepsilon_0 = 0$ and 8 T at the N = 0 LL ($\nu_{tot} = -1$ and $-3$), indicating compressible states. When increasing the magnetic field, the high resistance vanishes around 15 T, yielding a stable layer-coherent quantum Hall state. In contrast, the odd-integer quantum Hall states at N = 1 are already stable at 8 T; $\nu_{tot} = -5$ and $-7$ indicate resistivity minima or zero from 8 T, as shown in Figures S4d and S4e.

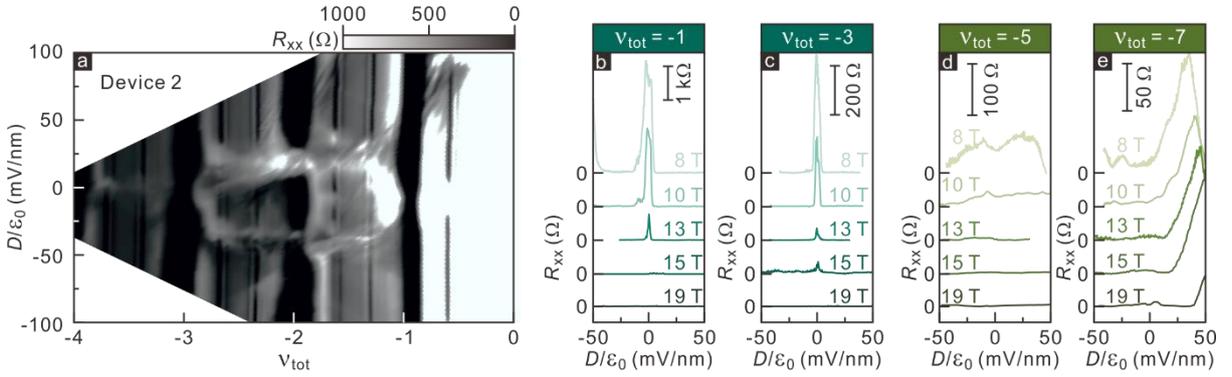

**Figure S4| Odd-integer quantum Hall states in the N = 0 LL.** (a) Greyscale plot of $R_{xx}$ in the window ($\nu_{tot}$, $D/\varepsilon_0$) for the N = 0 LL of Device 2 at $B$ = 19 T and $T$ = 30 mK. (b)–(d) $R_{xx}$ as function of $D/\varepsilon_0$ at $T$ = 30 mK for $\nu_{tot}$ = $-1$, $-3$, $-5$, and $-7$. Each quantum Hall state is labelled at the top of each panel. The five colours indicate the different magnetic fields. The five $R_{xx}$ curves are vertically offset for clarity. Scale bars indicate the longitudinal resistance for each panel.

There are several clear differences between small twist angle devices (~2°) and current large angle devices. More precisely, large angle device allows us to contrast the difference between the physics of N = 0 and N = 1 Landau level physics in the twisted bilayer graphene, which was missing in the small twist angle devices (~2°). Theoretically, exciton condensate can emerge not only at the N = 0 Landau level, but also at the higher levels. However, the exciton condensate, i.e. the interlayer-coherent quantum Hall state, appears only in the N = 0 Landau level on GaAs and (non-twisted) graphene double-layer systems. Hence, one may wonder if the same occurs in the twisted bilayer graphene systems.

To answer to this question, one obviously needs a system, which can access the N = 0 and N = 1 Landau levels in the same device. This could have not done in the 2° devices in our previous study. Since, the magnetic breakdown due to the van Hove singularity kicked in before getting to the N = 1 Landau level as shown in Fig S5. In this regime, the Fermi surfaces from the upper layer and down layer are interconnected, and the originally expected odd-integer quantum Hall states are completely washed out. Therefore, the question on the fate of the odd-integer quantum Hall states in higher Landau levels was a standing puzzle in Ref. 45.

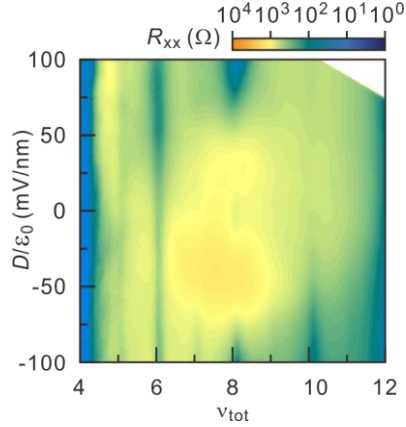

**Fig S5**| Color rendition of $R_{xx}$ in the $D/\varepsilon_0$ and $\nu_{tot}$ plane at $B = 6$ T, $T = 1.3$ K on the 2° twist device

This is exactly what we have achieved in our current devices. Naively one expects that the lower the level is, the more stable the exciton condensate is. However, we find that they are (surprisingly) absent in the N = 0 Landau level but present in the N = 1 Landau level. Theoretically this is attributed to the smaller energy scale than the electron-hole excitations, i.e. the gap for Skyrmion excitations, as discussed by Ref. 53. In contrast to this, the odd integer quantum Hall states, which are stabilized by the exciton condensation, are very robust in the N = 1 Landau level. Note that in the 2° devices, the odd integer quantum Hall states survived. However, they do have the smaller energy gaps than those in the N = 1 Landau levels in our current devices, which is consistent with our discussion here.

In summary, for the first time we discovered stable exciton condensation in the N = 1 Landau level, which is new to the twisted bilayer graphene community. This was enabled by the large-angle twist between the graphene layers. When combined with our previous study, our work clearly demonstrates that the fate of the odd integer quantum Hall states sensitively depends on the twist angle, particularly in the N = 1 Landau level.

## II. Theoretical model

In this section, we analyze the theoretical model used in our paper. To properly capture the phenomenology of the quantum Hall effects of twisted bilayer graphene at a large twist angle, we introduce the following model, which is a generalization of previous theoretical models [3–6].

The effective Hamiltonian projected onto the N = n LL is given by

$$
\hat{H}_{eff} = \int d^2x\,\hat{\psi}^\dagger(\boldsymbol{x})\left[\epsilon_n - \frac{\epsilon_Z}{2}\sigma_s^z + \frac{u}{2}\sigma_l^z\right]\hat{\psi}(\boldsymbol{x}) + \frac{\alpha}{4}\frac{d}{l_B}\int d^2x\,(n_t(\boldsymbol{x}) - n_b(\boldsymbol{x}))^2
$$
$$
+ \frac{1}{2}\int d^2x\,d^2x'\,V_C(\boldsymbol{x}-\boldsymbol{x}') : \hat{\psi}^\dagger(\boldsymbol{x})\hat{\psi}(\boldsymbol{x})\hat{\psi}^\dagger(\boldsymbol{x}')\hat{\psi}(\boldsymbol{x}'):
$$
$$
+ \int d^2x\,\, 2\pi l_B^2 \left[\sum_{l=\pm 1}\left(V_1 n_{l,K} n_{l,K'} - J_H \vec{S}_{l,K}\cdot\vec{S}_{l,K'}\right) + V_2\sum_{\xi=\pm}n_{t,\xi K}n_{b,\xi K} + V_3\sum_{l=\pm 1}n_{l,K}n_{-l,K'}\right], \qquad (1)
$$

where $\epsilon_n$ is the energy of the N = n LL, $\kappa$ is the effective dielectric constant of the graphene, $\alpha = \frac{e^2}{\kappa\,l_B}$, $l_B = \sqrt{\frac{\hbar c}{eB}}$ is the magnetic length, $u$ is the (displacement) electric field strength, $V_C$ is the Coulomb potential, $\hat{\psi}(\boldsymbol{x})$ is the LL projected electron annihilation operator, $n_{l,\xi K} = \sum_{s=\uparrow,\downarrow}\hat{\psi}^\dagger_{l,\xi K,s}(\boldsymbol{x})\,\hat{\psi}_{l,\xi K,s}(\boldsymbol{x})$ with $\xi = \pm$ and $-K = K'$, $n_l = \sum_{\xi=\pm}n_{l,\xi K}$ for the layer index $l = +1(-1)$ denoting the top (bottom) layer, $\left[\vec{S}_{l,\xi K}\right]_{s,s'} = \hat{\psi}^\dagger_{l,\xi K,s}(\boldsymbol{x})\,[\vec{\sigma}]_{s,s'}\,\hat{\psi}_{l,\xi K,s'}(\boldsymbol{x})$, and $:\cdots:$ denotes the normal ordering of field operators.

As we are interested in the physics at the fixed Landau level index N = 1 LL, the energy $\epsilon_n$ and the long-range part of the Coulomb interaction $V_C$ can be omitted for the following as they merely give constant energy shifts. The interaction effect is thus captured via short-range interactions, the third line of Eq. (1), where part of its coupling constant derives from the short-range part of the Coulomb interaction [5]. However, as other effects such as electron–phonon interactions do contribute to these short-range interactions, we treat the coupling constants of the short-range interactions as phenomenological parameters [5,7]. Moreover, we expect that the coupling constants depend on the LL index, similar to how the LL projected Coulomb interaction depends on the LL index [8]. With these assumptions and using the Fierz identities [6], we can further reduce the Hamiltonian as

$$
\hat{H}_{eff} = \int d^2x\,\hat{\psi}^\dagger(\boldsymbol{x})\left[-\frac{\epsilon_Z}{2}\sigma_s^z + \frac{u}{2}\sigma_l^z\right]\hat{\psi}(\boldsymbol{x}) + \frac{\alpha}{4}\frac{d}{l_B}\int d^2x\,(n_t(\boldsymbol{x}) - n_b(\boldsymbol{x}))^2
$$
$$
+ \sum_a \int d^2x\,\,2\pi l_B^2\,u_a : \left[\hat{\psi}^\dagger(\boldsymbol{x})\Gamma_a\hat{\psi}(\boldsymbol{x})\right]^2:, \qquad (2)
$$

where $a = (\mu,\nu)$ with $\mu,\nu = 0,x,y,z$, $\Gamma_{a=(\mu,\nu)} = \sigma_l^\mu\otimes\sigma_V^\nu$ with $l$ and $V$ representing the layer and valley space, and $u_{00} = \frac{1}{8}(-V_1 - J_H + V_2 + V_3)$, $u_{z0} = \frac{1}{8}(-V_1 - J_H - V_2 - V_3)$, $u_{0z} = \frac{1}{8}(V_1 + J_H + V_2 - V_3)$, $u_{zz} = \frac{1}{8}(V_1 + J_H - V_2 - V_3)$, $u_{0x} = u_{0y} = u_{zx} = u_{zy} = -\frac{1}{4}J_H$, and all other coefficients are 0.

## S3. The Hartree–Fock method

Having constructed the effective model, we now discuss the numerical method by which it was solved. Here, we employ the Hartree–Fock mean-field approach, which appropriately captures the physics of the integer quantum Hall effects in twisted bilayer graphene [3,4,7]. The Hartree–Fock method involves approximating the electronic ground state as a Slater determinant state using numerics. Assuming translation invariance, the Hartree–Fock mean-field solution can be represented using $P = \sum_{a=1}^{\tilde{\nu}} \chi_a \chi_a^\dagger$, where $\tilde{\nu}$ is the filling factor (related to the usual filling factor $\nu_{tot}$ via $\tilde{\nu} = \nu_{tot} - 4$ for N=1 LL) and $\chi_a$ is a set of orthonormal vectors in $\mathbb{C}^8$. The energy functional is given by

$$
\begin{aligned}
E_{MF}[P] = &-\frac{\epsilon_Z}{2} tr(P\sigma_s^z) + \frac{u}{2} tr(P\sigma_l^z) + \frac{\alpha}{4}\frac{d}{l_B}[tr(P\sigma_l^z)]^2 + \frac{1}{2}u_{00}[tr(P)^2 - tr(P^2)] \\
&+ \frac{1}{2}u_{z0}[tr(P\sigma_l^z)^2 - tr(P\sigma_l^z P\sigma_l^z)] + \frac{1}{2}u_{0z}[tr(P\sigma_V^z)^2 - tr(P\sigma_V^z P\sigma_V^z)] \\
&+ \frac{1}{2}u_{zz}[tr(P\sigma_l^z\sigma_V^z)^2 - tr(P\sigma_l^z\sigma_V^z P\sigma_l^z\sigma_V^z)] \\
&+ \frac{1}{2}u_{0x}[tr(P\sigma_V^x)^2 - tr(P\sigma_V^x P\sigma_V^x)] + \frac{1}{2}u_{0y}[tr(P\sigma_V^y)^2 - tr(P\sigma_V^y P\sigma_V^y)] \\
&+ \frac{1}{2}u_{zx}[tr(P\sigma_l^z\sigma_V^x)^2 - tr(P\sigma_l^z\sigma_V^x P\sigma_l^z\sigma_V^x)] \\
&+ \frac{1}{2}u_{zy}[tr(P\sigma_l^z\sigma_V^y)^2 - tr(P\sigma_l^z\sigma_V^y P\sigma_l^z\sigma_V^y)]
\end{aligned}
\tag{3}
$$

Note that the Hartree–Fock mean-field solution minimizes $E_{MF}[P]$. Since we fix the filling factor $\tilde{\nu}$ by constraining $tr(P) = \tilde{\nu}$ (and note that $P^2 = P$) throughout the computations, the term proportional to $u_{00}$ does not change. The corresponding mean-field Hamiltonian is then given by

$$
\begin{aligned}
H_{MF}[P] = &-\frac{\epsilon_Z}{2}\sigma_s^z + \frac{u}{2}\sigma_l^z + \frac{\alpha}{4}\frac{d}{l_B}tr(P\sigma_l^z)\sigma_l^z + u_{z0}[tr(P\sigma_l^z)\sigma_l^z - \sigma_l^z P\sigma_l^z] \\
&+ u_{0z}[tr(P\sigma_V^z)\sigma_V^z - \sigma_V^z P\sigma_V^z] + u_{zz}[tr(P\sigma_l^z\sigma_V^z)\sigma_l^z\sigma_V^z - \sigma_l^z\sigma_V^z P\sigma_l^z\sigma_V^z] \\
&+ u_{0x}[tr(P\sigma_V^x)\sigma_V^x - \sigma_V^x P\sigma_V^x] + u_{0y}[tr(P\sigma_V^y)\sigma_V^y - \sigma_V^y P\sigma_V^y] \\
&+ u_{zx}[tr(P\sigma_l^z\sigma_V^x)\sigma_l^z\sigma_V^x - \sigma_l^z\sigma_V^x P\sigma_l^z\sigma_V^x)] \\
&+ u_{zy}[tr(P\sigma_l^z\sigma_V^y)\sigma_l^z\sigma_V^y - \sigma_l^z\sigma_V^y P\sigma_l^z\sigma_V^y]
\end{aligned}
\tag{4}
$$

## S4. Mean-field solutions

To solve the mean-field Hamiltonian Eq. (4), we need to choose physical parameters that appear in the Hamiltonian. To this end, we chose the phenomenological values for the parameters $(J_H, V_1, V_2, V_3)$ guided by literature [2,4,5,7]. Specifically, the values of $J_H$ and $V_1$ are based on the values used to explain various phases in bilayer graphene [4,7]. As both $J_H$ and $V_1$ are responsible for intralayer couplings, we assume that their values are similar for our large-angle twisted bilayer graphene. As $V_2$ and $V_3$ are responsible for coupling across the layers, their strength would be smaller than the intralayer coupling strength $V_1$. Hence, the values for $V_2$ and $V_3$ are chosen such that they are smaller than $V_1$ and satisfy the relationship $|V_2| > |V_3|$ [4]. The model parameter values are chosen so that they agree with the quantum Hall gaps observed experimentally and the positions of the phase transitions in the displacement field $u$. The relative ratio between the parameters are guided by those of the corresponding Coulomb interactions. For instance, the ratio between the intralayer intervalley density-density interaction $V_1$ and the interlayer intravalley density-density interaction $V_2$ is similar to the ratio between the intralayer Coulomb interaction (which behaves as $\propto 1/|q|$ in 2D with the momentum $q$ being the distance between the valleys in momentum space) and the interlayer Coulomb interaction (which behaves as $\propto e^{-qd}/|q|$ with the momentum $q$ being the distance between the interlayer valleys in momentum space and $d$ being the interlayer spacing). Their precise values are different from the exact Coulomb interactions and used as the fitting parameters, because they are presumably strongly renormalized from their bare values and are phenomenological parameters in our model and [4, 6]. The values are summarized in Table S1. We have confirmed that the phases and phase diagram are stable with minor changes to the model parameters.

| Parameter | Value |
|---|---|
| Layer spacing d | $0.34$ nm |
| Magnetic length $l_B$ | $26/\sqrt{B/T}$ nm |
| $\alpha = \dfrac{e^2}{\kappa\, l_B}$ | $11.25/\sqrt{B/T}$ meV |
| Zeeman coupling $\epsilon_Z$ | $0.11\, B/T$ meV |
| $\alpha \dfrac{d}{l_B}$ | $0.15\, B/T$ meV |
| $J_H$ | $0.8 \times \alpha \dfrac{d}{l_B}$ |
| $V_1$ | $0.8 \times \alpha \dfrac{d}{l_B}$ |
| $V_2$ | $0.4 \times \alpha \dfrac{d}{l_B}$ |
| $V_3$ | $0.2 \times \alpha \dfrac{d}{l_B}$ |

**Table S1| Values of parameters used in the mean-field calculations.** Here, $B$ denotes the strength of the magnetic field. Note that paramters $J_H$, $V_1$, $V_2$, and $V_3$ increase linearly with $B$, as can be seen in References [4,7].

In this final section, we present the solutions of our mean-field equation. Note that our mean-field Hamiltonian features symmetries, including rotation in the valley space $e^{i\theta\sigma_V^z}$ and valley-exchange symmetry $\sigma_V^x$. To keep our presentation succinct, we only show the simplest solutions below, and other symmetry-related solutions are equally valid. The solutions are obtained via numerical minimization, since the analytical solutions are often not available. Moreover, we often get solutions with nonzero-layer coherence, particularly when $\nu_{tot}$ is odd and $u$ is small. This shows the existence of stable interlayer-coherent quantum Hall phases at odd $\nu_{tot}$ and with small electric fields. In this case, a single-particle eigenstate is supported on the top and bottom layers, and the amplitude changes smoothly as a function of the electric field.

In the following, we present the solutions where the electric field strength $u$ is greater than or equal to 0. The full phase diagram can be found in Figure 3c in the main text. For a fixed $\nu_{tot}$, the solutions are presented in the order by which they are realized when increasing $u$ from 0. The solution with $u \leq 0$ can be obtained by exchanging the layer index $t \leftrightarrow b$. Finally, the layer polarisation is defined as $tr(P\sigma_l^z)$.

## A. $\nu_{tot} = 5$

*Layer-coherent phase*: The mean-field solution has a non-zero layer off-diagonal coherence with ferromagnetic ordering owing to the Zeeman coupling. The layer polarization continuously varies as a function of $u$.

*Layer-polarized ferromagnetic phase*: The mean-field solution is given by $|\chi_1\rangle = |b, K, \uparrow\rangle$. This phase is characterised by a full layer polarization ($tr(P\sigma_l^z) = -1$) and has a ferromagnetic ordering.

## B. $\nu_{tot} = 6$

*Layer-coherent phase*: The mean-field solution has a non-zero layer off-diagonal coherence with ferromagnetic ordering owing to the Zeeman coupling. The layer polarization continuously varies as a function of $u$.

*Layer-polarized canted antiferromagnetic phase*: The solution is given by $|\chi_1\rangle = |b, K, \boldsymbol{s}(\theta)\rangle$ and $|\chi_2\rangle = |b, K', \boldsymbol{s}(-\theta)\rangle$, where the angle $\theta \geq 0$ satisfies $cos(\theta) = \frac{\epsilon_Z}{2J_H}$, $|\boldsymbol{s}(\theta)\rangle = cos(\frac{\theta}{2})|\uparrow\rangle + sin(\frac{\theta}{2})|\downarrow\rangle$, and the mean-field energy is given by $E_{MF} = \alpha \frac{d}{l_B} - u - \frac{V_1}{2} + \frac{V_3}{2} - \frac{J_H}{2} - \frac{\epsilon_Z^2}{4J_H}$. This phase has layer polarization $-2$.

## C. $\nu_{tot} = 7$

*Layer-coherent phase*: The mean-field solution has a non-zero layer off-diagonal coherence with ferromagnetic ordering owing to the Zeeman coupling. The layer polarization continuously varies as a function of $u$.

*Fully layer-polarized phase*: The mean-field solution is given by $|\chi_1\rangle = |b, K, \uparrow\rangle$, $|\chi_2\rangle = |b, K', \uparrow\rangle$, and $|\chi_3\rangle = |b, K, \downarrow\rangle$. The mean-field energy is given by $E_{MF} = \frac{9\alpha}{4} \frac{d}{l_B} - \frac{3u}{2} - V_1 + \frac{V_3}{2} - \frac{\epsilon_Z}{2}$. This phase has layer polarization $-3$.

## D. $\nu_{tot} = 8$

*Canted anti-ferromagnetic phase*: The mean-field solution is given by $|\chi_1\rangle = |t, K, \boldsymbol{s}(\theta)\rangle$, $|\chi_2\rangle = |t, K', \boldsymbol{s}(-\theta)\rangle$, $|\chi_3\rangle = |b, K, \boldsymbol{s}(\theta)\rangle$, and $|\chi_2\rangle = |b, K', \boldsymbol{s}(-\theta)\rangle$. This phase is a copy of

the canted antiferromagnetic phase with the mean-field energy $E_{MF} = -\frac{V_1}{2} + V_2 + \frac{3V_3}{2} - J_H - \frac{\epsilon_Z^2}{2J_H}$ with no layer polarization.

*Partially layer-polarized phase with layer coherence*: The mean-field solution has a layer coherence in this phase. While $|\chi_1\rangle = |b, K, \uparrow\rangle$ and $|\chi_2\rangle = |b, K', \uparrow\rangle$ have well-defined layer and valley indices, $|\chi_3\rangle$ and $|\chi_4\rangle$ are in superposition states over different layers. Owing to interlayer coherent states $|\chi_3\rangle$ and $|\chi_4\rangle$, the mean-field solution has continuously varying layer polarization. The solution also has a nonzero ferromagnetic order.

*Fully layer-polarized phase*: The mean-field solution is given by $|\chi_1\rangle = |b, K, \uparrow\rangle$, $|\chi_2\rangle = |b, K', \uparrow\rangle$, and $|\chi_3\rangle = |b, K, \downarrow\rangle$, and $|\chi_4\rangle = |b, K', \downarrow\rangle$. The mean-field energy is given by $E_{MF} = 4\alpha\frac{d}{l_B} - 2u - 2V_1 + \frac{V_3}{2}$ and the layer polarization is $-4$ in this case.

## E. $\nu_{tot} = 9$

*Layer-coherent phase*: The mean-field solution has a nonzero layer off-diagonal coherence with ferromagnetic ordering owing to the Zeeman coupling. The layer polarization continuously varies as a function of $u$.

*Layer-polarized phase*: The mean-field solution is given by $|\chi_1\rangle = |b, K, \uparrow\rangle$, $|\chi_2\rangle = |b, K', \uparrow\rangle$, $|\chi_3\rangle = |b, K, \downarrow\rangle$, $|\chi_4\rangle = |b, K', \downarrow\rangle$, and $|\chi_5\rangle = |t, K, \uparrow\rangle$. The corresponding mean-field energy is given by $E_{MF} = \frac{9\alpha}{4}\frac{d}{l_B} - \frac{3u}{2} - 2V_1 + V_2 + \frac{3V_3}{2} - \frac{\epsilon_Z}{2}$ and the layer polarization equals $-3$.

## F. $\nu_{tot} = 10$

*Layer-coherent phase*: The mean-field solution has a nonzero layer off-diagonal coherence with a ferromagnetic ordering owing to the Zeeman coupling. The layer polarization continuously varies as a function of $u$.

*Fully layer-polarized phase*: The mean-field solution is given by $|\chi_1\rangle = |b, K, \uparrow\rangle$, $|\chi_2\rangle = |b, K', \uparrow\rangle$, and $|\chi_3\rangle = |b, K, \downarrow\rangle$, $|\chi_4\rangle = |b, K', \downarrow\rangle$, $|\chi_5\rangle = |t, K, \mathbf{s}(\theta)\rangle$, and $|\chi_6\rangle = |t, K', \mathbf{s}(-\theta)\rangle$. The corresponding mean-field energy is given by $E_{MF} = \alpha\frac{d}{l_B} - u - \frac{5V_1}{2} + 2V_2 + \frac{11V_3}{4} - \frac{J_H}{2} - \frac{\epsilon_Z^2}{4J_H}$. This phase is characterised by a fully occupied bottom layer and canted antiferromagnetic phase in the top layer. The layer polarization equals $-2$ in this phase.

## G. $\nu_{tot} = 11$

*Layer-coherent phase*: The mean-field solution has a non-zero layer off-diagonal coherence with a ferromagnetic ordering owing to the Zeeman coupling. The layer polarization continuously varies as a function of $u$.

*Layer-polarized ferromagnetic phase*: The mean-field solution is given by $|\chi_1\rangle = |b, K, \uparrow\rangle$, $|\chi_2\rangle = |b, K', \uparrow\rangle$, $|\chi_3\rangle = |b, K, \downarrow\rangle$, $|\chi_4\rangle = |b, K', \downarrow\rangle$, $|\chi_5\rangle = |t, K, \uparrow\rangle$, $|\chi_6\rangle = |t, K', \uparrow\rangle$, and $|\chi_6\rangle = |t, K, \downarrow\rangle$. This phase is characterised by a layer polarization of $-1$ and has a ferromagnetic ordering.

In our calculation, we systematically found that at the vanishing displacement field, the energy gaps of odd fillings are smaller than that of even integer fillings. This can be attributed to the hierarchy of the interactions in the systems, where the intralayer interactions $J_H, V_1$ are larger than the interlayer couplings $V_2, V_3$. Furthermore, the gaps at the even fillings are mainly originated from the intralayer interaction $J_H, V_1$ and those at the odd fillings are mainly attributed to the interlayer interactions $V_2, V_3$. This can be seen from the following intuitions. In a simplified picture, the even integer quantum Hall states can be approximated by two decoupled integer quantum Hall states at zero displacement field. Using this picture, one can naturally expect that the gaps at the even fillings are dominated by intralayer coupling while the energy gap is dominated by interlayer coupling at odd fillings. This can be seen in the largest energy gap case, which is $\nu_{tot} = 8$, where the phase from theoretical calculation is indeed given by a two decoupled copy of canted antiferromagnetic phase of two graphene monolayers. At other even integer fillings, energy gap is smaller than that of $\nu_{tot} = 8$ due to the presence of interlayer coherence, but the energy gap is still larger than those in the odd fillings.

## S5. Skyrmion–anti-skyrmion excitations

In the main text, we presented electron–hole excitation gaps obtained from the Hartree–Fock calculations. In addition to the particle–hole excitations, there exist other types of excitations such as skyrmion and anti-skyrmion excitations. As explained in a theoretical work [9] and a recent paper [10], the excitation gap for the skyrmion–anti-skyrmion pair can be smaller than that of the particle–hole pair at the lowest Landau level. The excitation energy of the skyrmion–anti-skyrmion pair increases in the bilayer graphene while the excitation energy of the particle-hole pair decreases as a function of the Landau level index, so skyrmion–anti-skyrmion pairs can dominate low energy excitations at the N = 0 Landau level. Thus, skyrmion–anti-skyrmion pair excitations may explain the observations made in experiments where quantum Hall phases become more stable at the N = 1 Landau level than the N = 0 Landau level.

We employ skyrmion–anti-skyrmion pair excitations in untwisted bilayer graphene to estimate the excitation gap of the skyrmion–anti-skyrmion pair in our system. As the skyrmion–anti-skyrmion pair involves topologically nontrivial (pseudo)spin texture in real space, we believe the skyrmion–anti-skyrmion excitation energy in the usual untwisted bilayer system can be served as a zeroth order approximation of the skyrmion–anti-skyrmion excitation energy in our system. Using the result in Reference [9], the lowest excitation gap of the skyrmion–anti-skyrmion pair in bilayer graphene is given by $\Delta_{SK} = 8\pi\rho_s$, where $\rho_s$ is the stiffness given by

$$\rho_s = \frac{1}{32\pi^2} \int_0^\infty dq\, q^3\, V(q)\, [F_n(q)]^2 e^{-q^2 l_B^2/2} \tag{5}$$

with $V(q)$ being the Coulomb interaction in the momentum space and $F_n(q)$ is the structure factor of the N = n Landau level. In units of $\sqrt{\frac{\pi}{2}} \frac{e^2}{\kappa \, l_B}$, $\Delta_{SK}$ is equal to 0.5, 0.875, and 1.1328 when the Landau index N = 0, 1, and 2, respectively [7]. Although direct comparison between $\Delta_{SK}$ and electron–hole excitation gaps in our system is difficult, we note that $\Delta_{SK}$ is smallest at the N = 0 Landau level, and could be responsible for the observation that quantum Hall states are more stable in the N = 1 Landau level than in the N = 0 Landau level.

**References for Supporting Information**